\documentclass[preprint,12pt,authoryear]{elsarticle}
\journal{Journal of Systems and Software}
\usepackage[T1]{fontenc}
\usepackage{booktabs,tabularx,multirow,graphicx,amsmath,amssymb,microtype}
\usepackage[table]{xcolor}
\usepackage{xurl}
\usepackage[hidelinks]{hyperref}
\biboptions{round,semicolon}
\let\cite\citep
\providecommand{\Description}[2][]{}
\usepackage{tikz}
\usetikzlibrary{arrows.meta,positioning}
\definecolor{aegisTeal}{HTML}{176C83}
\definecolor{aegisRust}{HTML}{AA6035}
\definecolor{aegisInk}{HTML}{253347}
\definecolor{aegisPale}{HTML}{F0F6F8}
\definecolor{aegisPurple}{HTML}{8064A2}
\definecolor{aegisOlive}{HTML}{71883B}
\definecolor{aegisBlue}{HTML}{4477AA}
\newcommand{\method}{\textsc{Aegis}}
\newsavebox{\rqanswerbox}
\newenvironment{rqanswer}[1]{%
  \par\addvspace{0.7\baselineskip}%
  \begingroup
  \setlength{\fboxsep}{8pt}%
  \setlength{\fboxrule}{0.4pt}%
  \begin{lrbox}{\rqanswerbox}%
  \begin{minipage}{\dimexpr\linewidth-2\fboxsep-2\fboxrule\relax}%
  \noindent\textbf{Answer to #1.}\enspace\ignorespaces
}{%
  \end{minipage}%
  \end{lrbox}%
  \noindent\fcolorbox{black!35}{black!4}{\usebox{\rqanswerbox}}%
  \par\endgroup
  \addvspace{0.5\baselineskip}\ignorespacesafterend
}

\begin{document}
\raggedbottom
\begin{frontmatter}
\title{AEGIS: Evidence-Guided Set Decisions for Requirement-to-Code Traceability}
\author[osaka]{Zhiyao Wang}
\ead{wangzhiyao@ist.osaka-u.ac.jp}
\author[pku]{Yiming Peng}
\ead{pym@pku.edu.cn}
\author[dlmu]{Munan Li}
\ead{limunan@dlmu.edu.cn}
\author[swu]{Mingyue Zhang}
\ead{myzhangswu@swu.edu.cn}
\author[cqec]{Guanghong Li}
\ead{15902365958@163.com}
\author[waseda]{Jialong Li\corref{corresponding}}
\ead{lijialong@fuji.waseda.jp}
\author[osaka]{Tatsuhiro Tsuchiya}
\ead{t-tutiya@ist.osaka-u.ac.jp}
\cortext[corresponding]{Corresponding author.}
\affiliation[osaka]{organization={The University of Osaka},
  city={Suita}, country={Japan}}
\affiliation[pku]{organization={Peking University},
  city={Beijing}, country={China}}
\affiliation[dlmu]{organization={Dalian Maritime University},
  city={Dalian}, country={China}}
\affiliation[swu]{organization={Southwest University},
  city={Chongqing}, country={China}}
\affiliation[cqec]{organization={Chongqing Energy College},
  city={Chongqing}, country={China}}
\affiliation[waseda]{organization={Waseda Institute for Advanced Study, Waseda University},
  city={Tokyo}, country={Japan}}

\begin{abstract}
Requirement-to-code traceability recovery identifies code artifacts that
implement a requirement, but manual review costs limit how many candidates
developers can inspect. Code relations can reveal candidates missed by textual retrieval,
yet can also introduce code unrelated to the requirement. We propose
\method{}, which combines retrieval evidence and code relations to
construct and revise trace sets under a project-wide review budget. Grouped retrieval and semantic retrieval
supply candidates. A set objective combines relevance to the requirement
with structural compatibility. Local search constructs sets of different
sizes, and dynamic programming selects one set per requirement within
the budget. Call-context admission then compares replacements using
evidence from arguments and return values in calls involving protected members,
while bounding retrieval-score loss. Across 13 projects, \method{} achieves
the highest project-macro average on nine of ten evaluation metrics among
the compared methods. Relative to same-corpus adaptations of TAROT-VSM
and FTLR+MC, it improves F$_1$@5 by 33.21\% and 52.28\%, and MAP by
38.58\% and 49.08\%, respectively.
\end{abstract}

\end{frontmatter}

\section{Introduction}
\label{sec:draft-introduction}
Requirement-to-code traceability link recovery (TLR) identifies connections
between requirements and the code artifacts that implement them, helping
developers locate and understand the implementation of required
behavior~\cite{antoniol2002recovering,clelandhuang2014trends}. Controlled
maintenance experiments have shown that access to requirements traces can
help developers complete tasks more quickly and accurately~\cite{mader2015benefit}.
Retrieval-based approaches suggest candidate links for developers to
review~\cite{hayes2006candidate}. Developers check whether each suggested artifact contributes to implementing
the requirement.

Implementing a requirement may involve several code artifacts with different
responsibilities, and the number involved can vary across requirements.
Account creation, for example, may require a controller,
a validation component, and a persistence service. Their identifiers and
comments may describe what each component does without explicitly
mentioning account creation.
To find these artifacts, retrieval-based TLR uses requirement text as a
query and ranks code artifacts by their estimated relevance. Lexical
retrieval matches terms in requirements with those in code and identifiers,
while semantic retrieval can find related descriptions that use different
terms. Combining these retrieval methods can recover links missed by a
single method~\cite{gethers2011orthogonal}. Code relations can help identify
other artifacts that work together to implement the requirement, including
those that textual retrieval may overlook~\cite{panichella2013structural,kuang2019frugal}.

A controller may depend on both a validator that implements a required
check and a logging utility shared by many features. Whether either
artifact should have a trace link depends on its contribution to
implementing the requirement. Returning more candidates increases the
review workload, but returning too few may leave out part of the
implementation. Score-based structural enhancement uses code relations
to adjust retrieval scores and reorder candidate
links~\cite{panichella2013structural}. Under a limited review budget, the
additional candidates found through these relations compete with existing
candidates for the available positions. We use requirement relevance and
relations among candidates to construct the selected set. Once its output
count is fixed, admitting another candidate requires replacing a selected
member. We use retrieval evidence and call context to assess these revisions.

We propose \method{} (Anchored Evidence-Guided Inference of Sets), a
framework for selecting and revising candidate trace sets in
requirement-to-code traceability recovery. The name refers to a shield:
the leading members of the reference set remain protected during revision.
\method{} pools retrieval candidates and scores sets by requirement
relevance and structural compatibility. Dynamic programming selects one set for each requirement under the project
budget derived from retrieval scores. Directed code
relations then identify candidates for revision. Call-context admission
compares them with selected members using retrieval evidence and calls
involving the protected core. The final ranking places selected members first.

Our method constructs trace sets and bounds retrieval-score loss during revision.
We evaluate \method{} on 13 projects against 43 retrieval, fusion,
representation, propagation, and adapted traceability configurations,
together with same-graph diffusion.
With the same retrieval inputs and per-requirement output counts,
\method{} improves native F$_1$ by 6.14\% over the retrieval foundation,
with gains on 12 projects.
It also improves all ten project-macro recovery and ranking metrics over
this foundation.
Across the full comparison, \method{} achieves the highest project-macro
mean on nine of ten metrics. Relative improvements in
F$_1$@5 and MAP over the same-corpus TAROT-VSM adaptation are 33.21\% and
38.58\%, respectively; the corresponding improvements over the FTLR+MC
adaptation are 52.28\% and 49.08\%. Further comparisons examine the
contribution of code connections, changes to selected membership, and
recovery across five retrieval input pairs.

Sections~\ref{sec:draft-background} and~\ref{sec:draft-method} introduce
the problem and method; Sections~\ref{sec:draft-evaluation-design}
and~\ref{sec:draft-results} present the evaluation. Discussion, threats
to validity, related work, and conclusions follow in
Sections~\ref{sec:draft-discussion}--\ref{sec:draft-conclusion}.

\section{Background and Motivation}
\label{sec:draft-background}
Retrieval and code relations provide evidence for selecting artifacts
within a review budget. Once the set size is fixed, adding a candidate
requires choosing a member to replace.

\subsection{Retrieval Evidence and Set Selection}
\label{sec:draft-background-evidence}
\label{sec:draft-background-replacement}
Retrieval methods rank artifacts by lexical or semantic similarity for
analysts to review~\cite{antoniol2002recovering,borg2014mapping,hayes2006candidate}.
Combining views can find candidates missed by either
alone~\cite{gethers2011orthogonal}. Our code graph records calls, imports,
and inheritance to connect implementations and shared infrastructure
(Section~\ref{sec:draft-overview}).
\emph{Structural compatibility} measures how strongly these relations
support selecting artifacts together.

For account creation, a controller and a validator may form a useful
two-member set. A persistence component may justify a third position
if its requirement relevance and compatibility raise the set's score.
A shared utility may have weaker relevance or compatibility. If the
reference set contains the utility alongside the controller and validator,
a proposed persistence component may justify replacing the utility
while leaving the validator in place. We call
the proposed candidate the \emph{entrant} and the member it would
replace the \emph{donor}. Each replacement comparison considers the evidence supporting both
artifacts; arguments and returned values
in calls involving protected members provide further evidence of their roles.

\subsection{Problem Formulation}
\label{sec:draft-problem}
Let $Q$ denote a project's requirements, $C$ its code artifacts, and
$G$ a directed relation graph over $C$. Available evidence consists of
requirement-to-code retrieval scores, the relations in $G$, and source-call
fragments. For each $q\in Q$, the task returns a trace set $S_q^*\subseteq C$
and a complete ordering $\pi_q$ of $C$ whose first $|S_q^*|$ entries
contain exactly that set.

Let $\mathcal{B}$ be the total number of requirement--artifact links to
return for review. An artifact returned for two requirements occupies
two positions. \method{} derives this budget from retrieval evidence
and chooses reference sets $S_q^0$ with
$\sum_{q\in Q}|S_q^0|=\mathcal{B}$, using requirement relevance and
structural compatibility. Each set determines an output count
$k_q=|S_q^0|$, which remains fixed during revision. Replacements preserve
a protected core $H_q\subseteq S_q^0$ and produce a final set with
$|S_q^*|=k_q$ and $H_q\subseteq S_q^*$.

\section{The AEGIS Approach}
\label{sec:draft-method}
\method{} constructs reference sets under a project-wide review budget
and then assesses whether structural candidates should replace selected
members. The
implementation targets Java and uses retrieval scores, a code relation
graph, and source-call evidence.

\subsection{Overview}
\label{sec:draft-overview}
\begin{figure}[htbp]
\centering
\includegraphics[width=\linewidth]{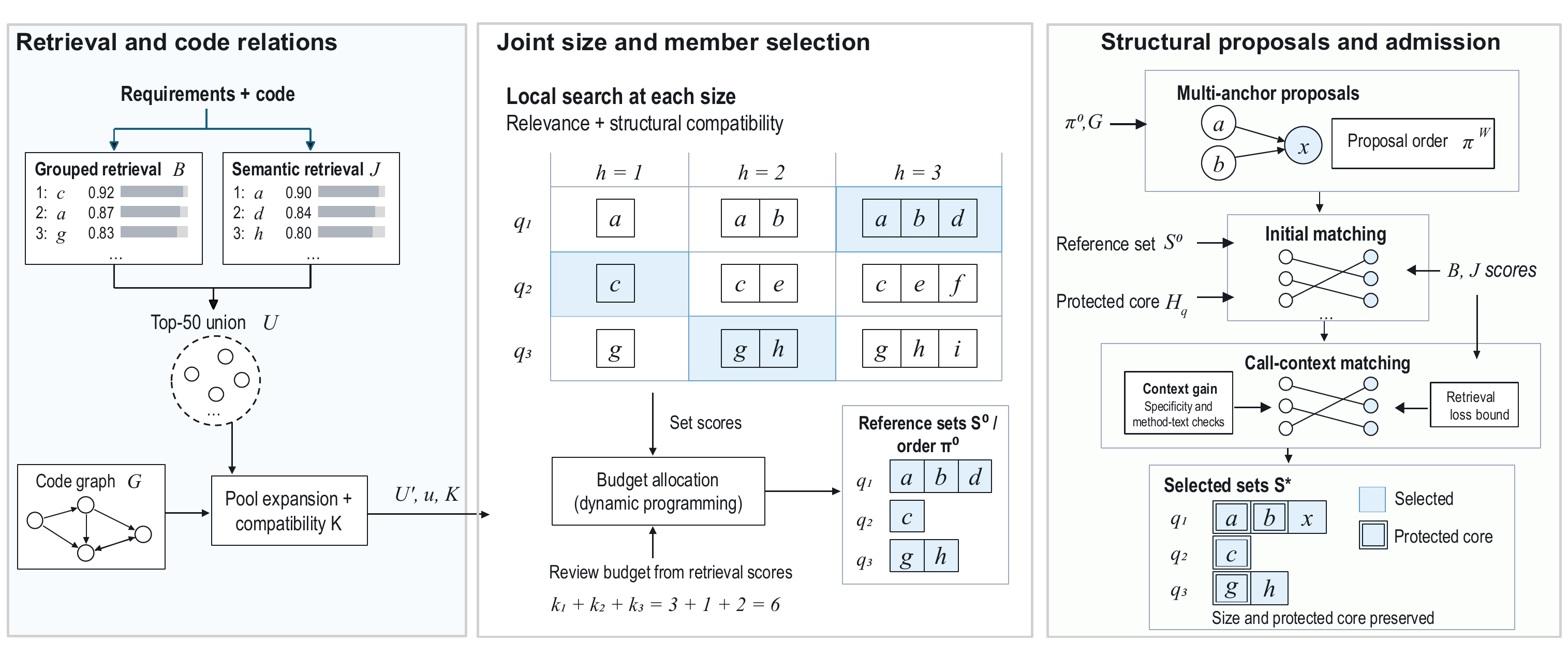}
\caption{AEGIS overview. Retrieval evidence and code relations support joint size and member selection, followed by structural proposals and two-stage admission. Admission preserves the allocated set sizes and protected cores.}
\label{fig:draft-framework}
\Description{A left-to-right workflow with three modules. Grouped retrieval
B and semantic retrieval J supply the top-50 union U. The code graph G
supports pool expansion and compatibility K. Local search constructs
candidate sets at different sizes using relevance and structural
compatibility. Dynamic programming allocates the total review budget
using the set scores. In the example, the reference sets contain a, b,
and d for q1; c for q2; and g and h for q3, with six positions in total.
The reference order and code graph produce multi-anchor proposals.
Initial matching uses the proposal order, reference set, protected core,
and B and J scores. Call-context matching uses context gain, specificity
and method-text checks, and the retrieval-loss bound. The final sets
replace d with x for q1 and leave q2 and q3 unchanged. Light-blue cells
mark selected members, and double borders mark protected members.}
\end{figure}

Figure~\ref{fig:draft-framework} shows the workflow. Retrieval views
and code relations supply candidates and structural compatibility.
\emph{Joint size and member selection} scores sets by requirement
relevance and structural compatibility, and allocates the total review
budget across requirements.
\emph{Structural proposal generation} uses directed connections to
multiple highly ranked artifacts to prioritize revision candidates.
\emph{Call-context admission} first matches candidates to unprotected
positions using retrieval and witness evidence, then compares alternatives
using calls involving the protected core. Replacements bound retrieval
loss relative to the original member; the final ranking places selected
members first. The requirement subscript is omitted where the context is clear.

\paragraph{Constructing the directed code relation graph.}
We construct $G=(C,\mathcal{E})$ over candidate Java files, identified by
unique class/file stems. In the edge set $\mathcal{E}$,
$c_i\rightarrow c_j$ means that $c_i$
depends on $c_j$. For call-based inputs, caller and callee classes give
the edge endpoints. For source-based inputs, imports and identifiers in
\texttt{extends} and \texttt{implements} clauses are matched to candidate
class names. We merge duplicate edges, exclude self-edges and external
references, and retain isolated candidates. The Boolean adjacency matrix
uses the same artifact order as the retrieval score matrices and is
shared across all requirements in the project. Proposals use directed reachability; reference compatibility uses the
symmetrized graph. Admission resolves
source calls separately to obtain argument and return fragments without
changing this graph.

\subsection{Joint Size and Member Selection}
\label{sec:draft-reference}
Reference construction jointly determines output counts and members
using requirement relevance and structural compatibility under the
project's total review budget. For each requirement, it returns
a reference set $S^0$ and a complete reference order $\pi^0$ for
later proposal generation and revision.

\paragraph{Retrieval evidence and total review budget.}
For each requirement, grouped evidence $B$ combines retrieval channels,
while semantic evidence $J$ compares requirement and code representations.
A channel pairs a scoring method with a requirement-text view and a
code-text view. Supplementary Section~S4.2 specifies the fields,
preprocessing, and fallback rules.

Channels form three groups: broad IR and graph evidence (18), focused
BM25 (4), and naming evidence (4). For a nonconstant score vector $v$,
let $\operatorname{rank}_v(c)$ be artifact $c$'s descending score rank.
Its log-rank score is
$L(v)(c)=1/\log_2(1+\operatorname{rank}_v(c))$.
Ranks start at 1, ties receive average ranks, and constant channels
contribute zero. Group means give $A$, $F$, and $N$, with
$B=(A+F+N)/3$. A shared channel contributes to each group containing it.

The channels include BM25 variants, character matching, LSI, LDA, and
graph propagation from source VSM scores~\cite{deerwester1990lsi,blei2003lda}.
Supplementary Sections~S4.1--S4.2 give their complete definitions.

The frozen
\path{jinaai/jina-code-embeddings-0.5b}
encoder~\cite{kryvosheieva2025jina} compares the full requirement with
the filename and source code. Cosine similarity of their L2-normalized
last-token representations gives $J$ (Supplementary Section~S4.2).

The initial pool $U$ is the union of candidates with average rank at
most 50 in either $B$ or $J$; constant views are omitted. We min--max
normalize each view on $U$, average the normalized scores with equal
weights, and min--max normalize the average on $U$ to obtain evidence
$E$. A normalization with an empty reference pool or zero score range
returns zero, and $E(c)=0$ outside $U$.

For each requirement $q$, we count candidates with positive evidence
$E_q(c)$ at or above $\tau_{\mathrm{out}}=0.60$. Summing these counts
gives the project's total review budget $\mathcal{B}$; an empty pool
contributes zero. Set selection allocates these positions across
requirements based on the scores of their candidate sets. Pool members
are initially ordered by the average normalized retrieval score,
followed by the remaining artifacts in $B$ order. Average-rank cutoffs
retain entire qualifying tie groups.

\paragraph{Structural compatibility and pool expansion.}
Reference selection considers both requirement relevance and relations
among artifacts. On the symmetrized code graph, direct and two-hop
connections provide pairwise support, with highly connected intermediate
nodes discounted by inverse degree. We subtract a degree-based
background, normalize by the geometric mean of the endpoint degrees,
and clip the result to $[-1,1]$ to obtain compatibility $K_{ij}$ between
artifacts $i$ and $j$. Diagonal entries and pairs involving an isolated
node receive zero. Positive compatibility rewards support above the
background, while negative compatibility penalizes support below it.

Structural support also extends the retrieval pool. Candidates with evidence above 0.5 serve as weighted seeds, whose support is
propagated over the same direct
and two-hop connections and discounted by the receiving candidate's
degree. We apply the same average-rank cutoff of 50 to the propagated
scores and add the qualifying candidates to obtain $U'$. For candidates in $U'$,
we apply the original pool's normalization scales to $B$ and $J$, average
the normalized views, normalize that average on $U$, and clip it to
$[0,1]$. The result is the per-artifact retrieval evidence $u$, also called unary
evidence. Keeping these scales fixed
preserves the evidence values of existing pool members as new candidates
enter the pool. Supplementary Section~S4.3 gives the normalization,
compatibility, and pool-expansion formulas.

\paragraph{Candidate sets and budget allocation.}
For a candidate set $S\subseteq U'$, let $h=|S|$. Its score is
\begin{equation}
 \Phi(S)=\sum_{c\in S}u(c)
 +\lambda_h\sum_{\{i,j\}\subseteq S}K_{ij},
 \qquad \lambda_h=\frac{1}{\sqrt{\max(h,1)}}.
 \label{eq:draft-reference-objective}
\end{equation}
The first term rewards requirement relevance, while the second measures
structural compatibility among the selected artifacts, counting each
unordered pair of distinct members once.

For each requirement $q$, we construct a set for every size
$h\in\{0,\ldots,m_q\}$, where $m_q=|U'_q|$ is its expanded-pool size.
For a given $h$, search starts from the top-$h$ unary candidates and
repeatedly takes the best improving one-member replacement.
Search stops when no gain exceeds the numerical tolerance
or after $\min(50,2|U'|)$ iterations. Write the resulting size-$h$ set as
$S_q^{(h)}$ and its score as $v_q(h)=\Phi_q(S_q^{(h)})$, where
$\Phi_q$ uses requirement $q$'s evidence. For $h=0$, the set is empty
and $v_q(0)=0$; an empty pool has this single option.

We choose one of these sets for each requirement by solving
\begin{equation}
 \max_{\{k_q\}_{q\in Q}}\ \sum_{q\in Q}v_q(k_q)
 \quad\text{subject to}\quad
 \sum_{q\in Q}k_q=\mathcal{B},\qquad
 k_q\in\{0,\ldots,m_q\}.
 \label{eq:draft-joint-allocation}
\end{equation}
Each possible output count is evaluated using the score of its candidate
set.

Dynamic programming solves Equation~\ref{eq:draft-joint-allocation}
over the sets produced by the local searches. Backtracking gives
$k_q$ and $S_q^0=S_q^{(k_q)}$ for every requirement. Each $k_q$ then
stays fixed during ordering and revision. Supplementary Section~S4.3
gives the exchange gains, recurrence, and tie rules.

\paragraph{Reference priority.}
Reference order $\pi^0$ supplies anchors and protected members.
We build this order by greedily adding candidates using unary evidence and
structural compatibility, with a cost for each artifact. We then append the
remaining pool candidates in order of marginal score, followed by candidates
outside the pool in their initial retrieval order. Finally, we move members
of $S^0$ to the front without changing the relative order within either
group. Supplementary Section~S4.3 gives
the ranking objective and procedure.

\subsection{Structural Proposal Generation}
\label{sec:draft-proposals}
We use directed connections to multiple anchors to order revision candidates
while keeping the reference set unchanged.

The first $\min(8,|C|)$ artifacts in $\pi^0$ serve as anchors, which
can extend beyond $S^0$ when $k<8$.

For $\ell\in\{1,2,3\}$, let $a_{\ell,\mathrm{out}}(c)$ count distinct anchors
reachable from $c$ by a directed path of at most $\ell$ edges, and
$a_{\ell,\mathrm{in}}(c)$ count distinct anchors that can reach $c$ within
the same bound. Self-reachability is excluded, and multiple paths to the
same anchor contribute one witness. We write $a_{\ell,\delta}(c)$ for either
direction, where $\delta\in\{\mathrm{out},\mathrm{in}\}$, and let $M(c)$ be
the number of anchors other than $c$ itself.
The witness score is
\begin{equation}
 W(c)=\max_{\ell,\delta}
 \frac{\mathbb{I}[a_{\ell,\delta}(c)\geq2]\,a_{\ell,\delta}(c)}
 {\sqrt{\ell}\,M(c)},
 \label{eq:draft-witness}
\end{equation}
with zero for a zero denominator. The indicator $\mathbb{I}[\cdot]$
equals 1 when its condition holds and 0 otherwise. Two distinct anchors are required;
the distance discount favors nearby support, and the maximum retains
the strongest direction and path-length bound.

To combine witness support with reference priority, we assign the
descending values of $E$ to artifacts in $\pi^0$ order without changing
the values. A reference-order tie-breaking term gives the assigned scores
$\widetilde E$ (Supplementary Section~S4.3). Sorting
$\widetilde E+0.5W$ yields proposal order $\pi^W$. When witness support
is zero, this order follows $\pi^0$. These assigned scores determine proposal priority; admission
continues to use the original candidate-level $B$ and $J$ scores.

\subsection{Call-context Admission}
\label{sec:draft-admission}
\method{} compares replacement candidates using evidence from source
calls to or from protected core members.

\paragraph{Protected core and initial matching.}
The first $\lceil k/2\rceil$ members of $\pi^0$ form the protected core
$H_q$. Each original member $b\in S^0\setminus H_q$ defines a replaceable
position. An initial matching assigns a member $d_b$ to each position;
unmatched positions retain $d_b=b$. Candidates must have two-anchor
support and appear before $b$ in the proposal order. The initial matching
combines retrieval-score change with calibrated witness-score change;
the matching weight must be positive.

Calibration compares each artifact's anchor support across the project's
requirements. We also compare the raw witness-score difference between a
candidate and member $b$ across requirements. Its upper-tail fraction must
be smaller than its lower-tail fraction for the pair to qualify.
Supplementary Section~S4.5 gives the
scoring formulas and matching rules.

We min--max normalize $B$ and $J$ using the first $\min(100,|C|)$ artifacts
in $\pi^0$, obtaining $\bar B_q$ and $\bar J_q$. The transforms apply to
all candidates without clipping; a zero score range within this prefix
maps the view to zero. Every non-reference choice $a$ must satisfy
\begin{equation}
 \Delta_{\mathrm{ret},q}(a,b)=
 \frac{\bar B_q(a)-\bar B_q(b)+\bar J_q(a)-\bar J_q(b)}{2}
 \geq-0.20.
 \label{eq:draft-loss-test}
\end{equation}
Here, $\Delta_{\mathrm{ret},q}$ is the mean retrieval-score change relative
to the original member $b$.

\paragraph{Call-context evidence.}
Calls between a candidate and $H_q$ supply argument expressions,
parameter operations, return values, and call-site uses. We score each fragment by its term overlap with the requirement, weighted by
inverse document frequency (IDF). We normalize
this overlap by the square root of the fragment's distinct-term count
and the total IDF weight of requirement terms in the vocabulary of
method names, method bodies, and call fragments. Empty fragments score
zero. We take the highest argument-expression score and the highest
parameter-operation score separately, then use their geometric mean as
argument support. Return support is the geometric mean of return-value and call-site use
scores. Let $e(t,c\mid H_q)$ be the largest argument or return support
across calls in either direction between $c$ and $H_q$, scored against
requirement $t\in Q$, or zero if none provides support.

\paragraph{Requirement-specific comparisons.}
For an alternative $a$ to the initial member $d_b$, the call-context
gain is $D_t(a,d_b\mid H_q)=e(t,a\mid H_q)-e(t,d_b\mid H_q)$.
We score the same pair against every project requirement, keeping
$H_q$ and the source calls fixed. Its upper-tail fraction is
\begin{equation}
 T_q(a,d_b)=\frac{1}{|Q|}\sum_{t\in Q}
 \mathbb{I}\!\left[D_t(a,d_b\mid H_q)\geq D_q(a,d_b\mid H_q)\right].
 \label{eq:draft-context-tail}
\end{equation}
A change requires $D_q(a,d_b\mid H_q)>0$ and $T_q(a,d_b)\leq\alpha$,
where $\alpha=0.05$.
No more than 5\% of project requirements may give the same pair an
advantage at least as large as that for $q$. This fraction includes $q$
and counts ties. Call-context gain is measured against $d_b$; the
retrieval-score bound remains relative to the original member $b$.

\paragraph{Assignment and output.}
Additional candidates must appear among the first 50 positions of $\pi^W$,
lie outside the reference and initial sets, and have two-anchor support.
A position may also restore its original member $b$. Each alternative
must satisfy Equation~\ref{eq:draft-loss-test} and have a method-text score
at least as high as that of $d_b$. This score is the maximum over the artifact's methods, each scored from its
name and body using the same term-matching measure.

Among alternatives that pass these tests,
maximum-weight matching maximizes the sum of positive call-context gains.
Keeping $d_b$ has weight zero. Initial members and restored original members can occupy only their
own positions; each additional candidate can occupy at most one position. The resulting set $S^*$ contains $k$ members,
including $H_q$.

The final ranking places selected members first and the remaining
candidates afterward, preserving proposal order within both groups.

\subsection{Algorithmic Properties}
\label{sec:draft-properties}
With fixed inputs and deterministic tie handling, dynamic programming
finds an optimal allocation over the sets produced by the size-specific
local searches. The total size is $\mathcal{B}$. Admission assigns one
distinct artifact to every unprotected position, preserving $|S^*|=k$
and $H_q\subseteq S^*$. Search has a finite iteration cap, and admission
uses two finite matchings. Final ordering changes no selected members.

Every final replacement satisfies
$\Delta_{\mathrm{ret},q}(a_b,b)\geq-\tau$, where $a_b$ is the final
artifact at original member $b$'s position and $\tau=0.20$. Retained
initial replacements inherit the bound, restored originals have zero
change, and new context-based replacements are checked against $b$.
The mean change over replaced positions is therefore at least $-\tau$.
Supplementary Section~S4.6 gives the property derivations and time and
space bounds. Section~\ref{sec:draft-results-robustness} reports execution times.

\section{Evaluation Design}
\label{sec:draft-evaluation-design}
We compare recovery at common inspection budgets, examine set construction
and revision through controlled comparisons, and vary retrieval inputs
and parameters. The supplement provides complete configurations and protocols.

\subsection{Research Questions}
\label{sec:draft-rqs}
We address three research questions:
\begin{description}
 \item[RQ1: Recovery effectiveness.] How effectively does \method{}
 recover trace links at common inspection budgets?
 \item[RQ2: Set construction and revision.] How do retrieval and structural
 evidence affect set construction, and how do ordering and admission
 change the resulting output?
 \item[RQ3: Retrieval inputs and efficiency.] How do retrieval inputs
 and parameter settings affect recovery, and what is the execution cost
 of constructing and revising sets?
\end{description}

\subsection{Datasets and Evaluation Population}
\label{sec:draft-datasets}
% Generated from sealed results; do not edit values manually.
\begin{table}[htbp]\centering\small
\caption{Characteristics of the 13 evaluation projects.}
\label{tab:datasets}
\begin{tabular}{@{}lrrrrrl@{}}\toprule
Project & Requirements & Code & Links & Empty & Missing & Graph \\ \midrule
albergate & 17 & 55 & 53 & 1 & 0 & C \\
eANCI & 139 & 55 & 567 & 100 & 0 & C \\
eTour & 58 & 116 & 308 & 1 & 0 & C \\
iTrust & 131 & 226 & 286 & 26 & 0 & C \\
smos & 67 & 98 & 1027 & 0 & 17 & C \\
Derby & 390 & 2200 & 2315 & 0 & 0 & P \\
Drools & 183 & 3530 & 841 & 0 & 0 & P \\
Maven & 36 & 786 & 151 & 0 & 0 & P \\
Pig & 87 & 1316 & 546 & 0 & 0 & P \\
Seam2 & 189 & 1156 & 462 & 0 & 0 & P \\
Infinispan & 232 & 4105 & 1116 & 0 & 0 & P \\
Groovy & 104 & 100 & 180 & 0 & 0 & P \\
Dronology & 58 & 184 & 393 & 1 & 0 & P \\
\midrule Total & 1691 & 13927 & 8245 & 129 & 17 & --- \\
\bottomrule\end{tabular}
% AEGIS table note begin
\par\smallskip
\begin{minipage}{\linewidth}\footnotesize\raggedright
Empty: requirements without aligned gold links; missing: gold pairs with unavailable code candidates. C: call-derived graph; P: import/inheritance proxy.
\end{minipage}
% AEGIS table note end
\end{table}

Table~\ref{tab:datasets} summarizes 13 Java projects with 1,691 requirements,
13,927 code candidates, and 8,245 aligned positive links. They cover
management, municipal services, tourism, healthcare, and software infrastructure.
albergate supplies requirement-to-class links; eANCI, eTour, iTrust, and
smos supply use-case-to-class links. Eight snapshots come from the
HGNNLink data distribution~\cite{hgnnlinkdata}: seven derive from
issue/change-to-class data described by TAROT~\cite{gao2022tarot}, and
Dronology's upstream dataset describes manually established requirement-to-code
relations~\cite{dronologydata,clelandhuang2018dronology}. We use each snapshot's supplied trace matrix
as the gold standard. Supplementary Section~S5.1 details provenance and alignment.

All methods share requirement identifiers, candidates, and annotations.
The primary population includes all 1,691 requirements, of which 129
have no aligned positive; the linked-requirement analysis uses the other
1,562. In smos, 17 annotated pairs refer to unavailable code and fall
outside the shared universe. Graphs are call-derived for albergate,
eANCI, eTour, iTrust, and smos, and import/inheritance proxies for the
other projects, following Section~\ref{sec:draft-overview}.

\subsection{Baselines and Controlled Comparators}
\label{sec:draft-comparators}
We compare 43 baseline configurations covering lexical retrieval
(VSM, BM25, character TF--IDF, LSI, and Jensen--Shannon), pretrained
representations (Jina and UniXcoder), fusion (linear, reciprocal-rank,
and log-rank), graph propagation, and traceability-specific
methods~\cite{salton1975vsm,robertson2009bm25,deerwester1990lsi,cormack2009rrf}.
TAROT, FTLR, and O-CSTI are adapted to the shared
corpus~\cite{gao2022tarot,hey2021ftlr,tracelabcdk}; asterisks identify
these adaptations. Supplementary Sections~S1 and S4.4 list configurations
and adaptation details. UniXcoder uses the Jina input strings with its
official encoder formatting and pooling~\cite{guo2022unixcoder}.

The three primary comparators are:
\begin{enumerate}
 \item \emph{Same-input foundation} retains all retrieval channels,
 weights, pooling, and \method{}'s per-requirement counts, and returns
 the foundation ranking at those counts. It omits structural member
 selection, witness ordering, and admission.
 \item \emph{Equal-weight hybrid} combines normalized enriched BM25 and
 Jina scores across the candidate universe.
 \item \emph{Same-graph diffusion} propagates squared evidence $E_q^2$
 on the row-normalized undirected shared graph, with weight 0.35 and
 at most 50 iterations.
\end{enumerate}

\subsection{Controlled Experiments}
\label{sec:draft-experiments}
Comparisons use the 13 projects, aligned precomputed inputs, and
common defaults; sensitivity analyses vary the specified parameters.
Prediction uses requirements and code; gold links are used for evaluation. Supplementary Sections~S3 and S5 specify
the controls, repeated runs, and parameter grids.

\paragraph{Construction and ordering.}
Output-count controls compare fixed, uniform, and permuted counts.
Uniform and permuted allocation preserve each project's total budget;
permutation also preserves its count distribution. We test both the
complete pipeline and fixed foundation rankings. Member-selection
controls hold each requirement's count fixed, comparing the foundation,
joint selection, and unary top-$k$.
Locking reference membership isolates witness ordering.

To test observed code connections, a degree-only compatibility variant
removes one- and two-hop support while retaining the degree-background
penalty. Inputs, pools, counts, and downstream procedures remain fixed;
reference sets and priorities are recomputed. The comparison uses all
five retrieval input pairs. Further analyses examine candidate coverage
and links missed by both textual views against a graph-empty control.

\paragraph{Admission.}
Controls share reference sets, proposal orders, counts, and retrieval
scores. They compare membership locking, rank-only admission, initial
matching alone, removal of the specificity gate or method-text guard,
and whole-method evidence in place of call fragments. The no-core
variant removes both protected positions and their anchored call context.
Matched-activity controls retain each requirement's final replacement
count while varying exchange choices (Supplementary Section~S3.1).

\paragraph{Retrieval inputs, parameters, and cost.}
Five pairs combine grouped retrieval or enriched BM25 with Jina or
UniXcoder, plus enriched BM25 with VSM. Each pipeline is compared with
its foundation at matched per-requirement counts. Graphs and constants
remain fixed; score-dependent stages are recomputed. For the five-input
summary, scores are averaged within projects across input pairs before
project-level aggregation and inference.

Sensitivity tests vary nine parameters, including a $3\times4$
core-fraction--loss-tolerance grid and five specificity thresholds,
giving 28 distinct configurations. We measure source-feature preparation
separately from admission, repeat admission three times per project,
and sum all-size construction, allocation, and reference-ordering times.
Timings start from cached retrieval or parsed-call inputs as specified
in Supplementary Section~S2.4; they exclude retrieval scoring, embedding
extraction, and source parsing.

\subsection{Metrics and Statistical Analysis}
\label{sec:draft-metrics}
\paragraph{Recovery and inspection order.}
For requirement $q$, let $Y_q\subseteq C$ be its gold-linked artifacts.
We report precision, recall, F$_1$, and F$_2$; F$_1$ weights precision
and recall equally, while F$_2$ gives greater weight to recall.
Evaluation uses prefixes at depths 5, 10, and 20, native sets at equal
project budgets, and internal controls at \method{}'s per-requirement
counts $k_q$. External baselines and diffusion use the per-requirement threshold
counts described in Section~\ref{sec:draft-reference}; their sum equals the budget allocated by
\method{}. The foundation and internal matched-count controls share
each allocated $k_q$.

P@1 records whether the first suggestion is a gold link, MRR measures
the position of the first gold link, MAP uses all gold-link positions,
and binary-relevance nDCG@20 weights earlier links more
heavily~\cite{jarvelin2002gain}. The ten-metric profile comprises P@5,
Recall@5, F$_1$@5, F$_2$@5, MAP, MRR, P@1, nDCG@20, Recall@20, and
F$_2$@20. AP uses the complete ranking and denominator $|Y_q|$.
Zero-denominator cases and requirements without aligned positives
contribute zero. Supplementary Section~S5.3 gives the formulas.

\paragraph{Consequences of revision.}
For reference and final sets of equal size, revision downside is
$L_q=\max\{0,F_{2,q}(S_q^0)-F_{2,q}(S_q^*)\}$.
Losses are not offset by gains on other requirements. We report mean
downside and final native F$_2$. Gold-link additions and removals are
$|Y_q\cap(S_q^*\setminus S_q^0)|$ and
$|Y_q\cap(S_q^0\setminus S_q^*)|$. Each final entrant is paired with the original member at its assigned
position; intermediate assignments are excluded.

\paragraph{Aggregation and inference.}
Metrics, including F-measures, are computed per requirement, averaged
within projects, then averaged with equal project weights. Relative
gains are the macro difference divided by the comparator's macro mean;
percentage-point differences are the macro difference multiplied by 100.

Projects are the paired units of inference. We use 95\% paired project
bootstrap intervals~\cite{efron1979bootstrap}, exact two-sided sign-flip
tests over $2^{13}$ assignments, and paired rank-biserial effect
sizes~\cite{kerby2014difference}. Primary comparisons also use hierarchical
bootstrap intervals that resample projects and then requirements.
Holm adjustment~\cite{holm1979sequential} applies within comparison
families; $p_H$ denotes adjusted values. Repeated runs and input conditions
are averaged within projects, retaining 13 units of inference.
Mechanism and replacement contrasts require a positive mean, a 95\%
interval wholly above zero, and $p_H\leq0.05$ for statistical support.
Outside the call-context ablation family, F$_2$ differences accompanying
downside comparisons are descriptive. Supplementary Section~S5.4
specifies families, resampling settings, and assumptions.

\section{Results}
\label{sec:draft-results}
We report overall recovery, the effects of set construction and revision,
and performance across retrieval inputs, parameters, and execution stages.

\subsection{RQ1: Recovery Effectiveness}
\label{sec:draft-results-effectiveness}
\method{} leads nine of ten project-macro measures, covering short
prefixes, complete rankings, and depth-20 coverage (Table~\ref{tab:main}).
Supplementary Sections~S1 and S2.1 report all baseline configurations,
project-level values, and adapted traceability variants.
\begin{table}[htbp]
\centering\footnotesize
\setlength{\tabcolsep}{3pt}
\renewcommand{\arraystretch}{1.10}
\caption{Recovery effectiveness and ranking quality across 13 projects.}
\label{tab:main}
\label{tab:ranking}
\begin{tabular}{@{}lrrrr@{}}\toprule
\multicolumn{5}{@{}l}{\textit{(a) Recovery at five candidates}} \\
Method & P@5 & R@5 & F$_1$@5 & F$_2$@5 \\ \midrule
\method{} & \textbf{0.2440} & \textbf{0.3625} & \textbf{0.2338} & \textbf{0.2774} \\
\addlinespace[2pt]
\multicolumn{5}{@{}l}{\itshape Primary controls} \\
Same-input foundation & 0.2347 & 0.3512 & 0.2255 & 0.2681 \\
BM25e + Jina, equal weight & 0.2239 & 0.3513 & 0.2203 & 0.2652 \\
Same-graph diffusion & 0.2300 & 0.3600 & 0.2278 & 0.2737 \\
\addlinespace[2pt]
\multicolumn{5}{@{}l}{\itshape Retrieval and representation baselines} \\
BM25 (source) & 0.1626 & 0.2654 & 0.1619 & 0.1978 \\
Jina-only & 0.1977 & 0.2778 & 0.1851 & 0.2160 \\
RRF-19 & 0.2048 & 0.2993 & 0.1946 & 0.2299 \\
UniXcoder-only & 0.1087 & 0.1619 & 0.1036 & 0.1236 \\
\addlinespace[2pt]
\multicolumn{5}{@{}l}{\itshape Published traceability methods, adapted} \\
FTLR+MC* & 0.1642 & 0.2355 & 0.1535 & 0.1809 \\
TAROT-VSM* & 0.1915 & 0.2622 & 0.1755 & 0.2039 \\
\bottomrule\end{tabular}

\smallskip
\begin{tabular}{@{}lrrrrrr@{}}\toprule
\multicolumn{7}{@{}l}{\textit{(b) Ranking quality and coverage at twenty}} \\
Method & MAP & MRR & P@1 & nDCG@20 & R@20 & F$_2$@20 \\ \midrule
\method{} & \textbf{0.3654} & \textbf{0.5009} & 0.3731 & \textbf{0.4505} & \textbf{0.5749} & \textbf{0.2676} \\
\addlinespace[2pt]
Same-input foundation & 0.3424 & 0.4867 & 0.3620 & 0.4251 & 0.5424 & 0.2454 \\
BM25e + Jina, equal weight & 0.3408 & 0.4960 & \textbf{0.3765} & 0.4201 & 0.5270 & 0.2357 \\
Same-graph diffusion & 0.3524 & 0.4921 & 0.3670 & 0.4380 & 0.5682 & 0.2568 \\
\addlinespace[2pt]
BM25 (source) & 0.2649 & 0.3960 & 0.2752 & 0.3363 & 0.4545 & 0.1979 \\
Jina-only & 0.2835 & 0.4178 & 0.3180 & 0.3531 & 0.4572 & 0.2089 \\
RRF-19 & 0.2960 & 0.4245 & 0.3001 & 0.3760 & 0.5077 & 0.2304 \\
UniXcoder-only & 0.1698 & 0.2652 & 0.1585 & 0.2235 & 0.3290 & 0.1467 \\
\addlinespace[2pt]
FTLR+MC* & 0.2451 & 0.3592 & 0.2524 & 0.3132 & 0.4284 & 0.1983 \\
TAROT-VSM* & 0.2637 & 0.3738 & 0.2612 & 0.3386 & 0.4724 & 0.2160 \\
\bottomrule\end{tabular}
\par\smallskip
\begin{minipage}{\linewidth}\footnotesize\raggedright
Values are project-macro means. Asterisks mark same-corpus adaptations;
boldface marks column maxima. BM25e uses enriched code documents.
\end{minipage}
\end{table}

\begin{figure}[htbp]
 \centering
 \input{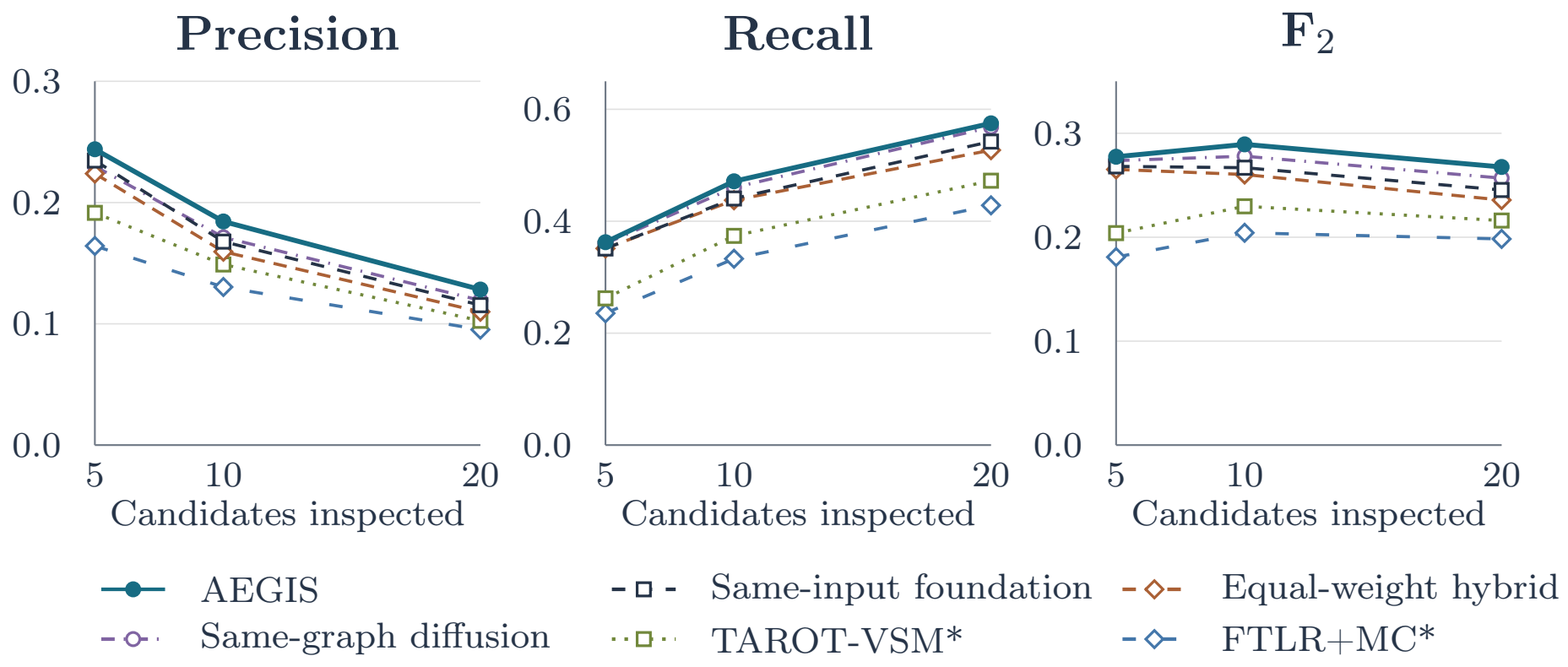}
 \caption{Precision, recall, and F$_2$ at inspection depths of 5, 10, and 20, macro-averaged across 13 projects.}
 \label{fig:budget-curves}
 \Description{Three line-chart panels plot precision, recall, and F2 against
 five, ten, and twenty inspected candidates for six fixed configurations.
 Every panel starts its vertical scale at zero. Straight segments connect
 the measurements at the three evaluated depths.}
\end{figure}

\paragraph{From retrieval evidence to recovered sets.}
\method{} has higher macro F$_1$@5 and MAP than the same-input
foundation, equal-weight hybrid, and same-graph diffusion controls
defined in Section~\ref{sec:draft-comparators}.
Figure~\ref{fig:budget-curves} shows how precision and coverage change
at the three evaluated inspection depths.

At the same total output count per project, \method{} improves precision,
recall, and both F-measures over each primary control
(Table~\ref{tab:native}). The foundation also shares \method{}'s
per-requirement counts. Against this control, native F$_1$ increases from
0.2537 to 0.2693, a relative gain of 6.14\%, with improvements on 12
projects and Holm-adjusted $p=8.30\times10^{-3}$. This comparison measures
changes in the selected membership at a fixed allocation.
\begin{table}[htbp]
\centering\small
\setlength{\tabcolsep}{3.5pt}
\renewcommand{\arraystretch}{1.10}
\caption{Recovery effectiveness under equal project review budgets.}
\label{tab:native}
\begin{tabular}{@{}lrrrr@{}}\toprule
Configuration & Precision & Recall & F$_1$ & F$_2$ \\
\midrule
\method{} & \textbf{0.2923} & \textbf{0.3820} & \textbf{0.2693} & \textbf{0.3045} \\
\addlinespace[2pt]
\multicolumn{5}{@{}l}{\footnotesize\itshape Controls under the same project budget} \\
Same-input foundation & 0.2757 & 0.3624 & 0.2537 & 0.2880 \\
BM25e + Jina, equal weight & 0.2788 & 0.3623 & 0.2528 & 0.2866 \\
Same-graph diffusion & 0.2847 & 0.3646 & 0.2581 & 0.2911 \\
\bottomrule\end{tabular}

% AEGIS table note begin
\par\smallskip
\begin{minipage}{\linewidth}\footnotesize\raggedright
All configurations use the same total output count per project. The foundation also uses AEGIS's count for each requirement. The hybrid and diffusion controls use shared per-requirement cutoffs from the retrieval-evidence threshold.
\end{minipage}
% AEGIS table note end
\end{table}

\begin{figure}[htbp]
 \centering
 \begin{tikzpicture}[x=1cm,y=1cm,every node/.style={font=\scriptsize,text=aegisInk}]
\node[anchor=east] at (2.82000,0.00000) {Same-input foundation};
\node[anchor=east] at (2.82000,-0.91000) {Equal-weight hybrid};
\node[anchor=east] at (2.82000,-1.82000) {Same-graph diffusion};
\node[font=\small\bfseries] at (4.86500,0.58000) {F$_1$@5};
\draw[black!12,line width=.3pt] (3.02000,0.24000) -- (3.02000,-2.19000);
\node[] at (3.02000,-2.45000) {-2};
\draw[black!12,line width=.3pt] (3.75800,0.24000) -- (3.75800,-2.19000);
\node[] at (3.75800,-2.45000) {+0};
\draw[black!12,line width=.3pt] (4.49600,0.24000) -- (4.49600,-2.19000);
\node[] at (4.49600,-2.45000) {+2};
\draw[black!12,line width=.3pt] (5.23400,0.24000) -- (5.23400,-2.19000);
\node[] at (5.23400,-2.45000) {+4};
\draw[black!12,line width=.3pt] (5.97200,0.24000) -- (5.97200,-2.19000);
\node[] at (5.97200,-2.45000) {+6};
\draw[black!12,line width=.3pt] (6.71000,0.24000) -- (6.71000,-2.19000);
\node[] at (6.71000,-2.45000) {+8};
\draw[aegisInk!70,line width=.6pt] (3.75800,0.24000) -- (3.75800,-2.19000);
\draw[aegisTeal,line width=1pt] (3.75417,0.08500) -- (4.32962,0.08500);
\draw[aegisTeal,line width=.6pt] (3.75417,0.04000) -- (3.75417,0.13000);
\draw[aegisTeal,line width=.6pt] (4.32962,0.04000) -- (4.32962,0.13000);
\path[draw=aegisTeal,fill=aegisTeal,line width=.6pt] (4.06499,0.08500) circle (0.052);
\draw[aegisInk!65,line width=1pt] (3.67280,-0.08500) -- (4.49987,-0.08500);
\draw[aegisInk!65,line width=.6pt] (3.67280,-0.13000) -- (3.67280,-0.04000);
\draw[aegisInk!65,line width=.6pt] (4.49987,-0.13000) -- (4.49987,-0.04000);
\path[draw=aegisInk!65,fill=white,line width=.6pt] (4.06499,-0.08500) circle (0.052);
\node[anchor=east] at (6.71000,-0.32000) {Holm $p=0.3601$};
\draw[aegisTeal,line width=1pt] (3.58065,-0.82500) -- (5.00447,-0.82500);
\draw[aegisTeal,line width=.6pt] (3.58065,-0.87000) -- (3.58065,-0.78000);
\draw[aegisTeal,line width=.6pt] (5.00447,-0.87000) -- (5.00447,-0.78000);
\path[draw=aegisTeal,fill=aegisTeal,line width=.6pt] (4.25376,-0.82500) circle (0.052);
\draw[aegisInk!65,line width=1pt] (3.48948,-0.99500) -- (5.10518,-0.99500);
\draw[aegisInk!65,line width=.6pt] (3.48948,-1.04000) -- (3.48948,-0.95000);
\draw[aegisInk!65,line width=.6pt] (5.10518,-1.04000) -- (5.10518,-0.95000);
\path[draw=aegisInk!65,fill=white,line width=.6pt] (4.25376,-0.99500) circle (0.052);
\node[anchor=east] at (6.71000,-1.23000) {Holm $p=0.6914$};
\draw[aegisTeal,line width=1pt] (3.50343,-1.73500) -- (4.67706,-1.73500);
\draw[aegisTeal,line width=.6pt] (3.50343,-1.78000) -- (3.50343,-1.69000);
\draw[aegisTeal,line width=.6pt] (4.67706,-1.78000) -- (4.67706,-1.69000);
\path[draw=aegisTeal,fill=aegisTeal,line width=.6pt] (3.97833,-1.73500) circle (0.052);
\draw[aegisInk!65,line width=1pt] (3.44991,-1.90500) -- (4.71032,-1.90500);
\draw[aegisInk!65,line width=.6pt] (3.44991,-1.95000) -- (3.44991,-1.86000);
\draw[aegisInk!65,line width=.6pt] (4.71032,-1.95000) -- (4.71032,-1.86000);
\path[draw=aegisInk!65,fill=white,line width=.6pt] (3.97833,-1.90500) circle (0.052);
\node[anchor=east] at (6.71000,-2.14000) {Holm $p=0.8799$};
\node[] at (4.86500,-2.80000) {AEGIS minus comparator (pp)};
\node[font=\small\bfseries] at (9.29500,0.58000) {MAP};
\draw[black!12,line width=.3pt] (7.45000,0.24000) -- (7.45000,-2.19000);
\node[] at (7.45000,-2.45000) {-2};
\draw[black!12,line width=.3pt] (8.18800,0.24000) -- (8.18800,-2.19000);
\node[] at (8.18800,-2.45000) {+0};
\draw[black!12,line width=.3pt] (8.92600,0.24000) -- (8.92600,-2.19000);
\node[] at (8.92600,-2.45000) {+2};
\draw[black!12,line width=.3pt] (9.66400,0.24000) -- (9.66400,-2.19000);
\node[] at (9.66400,-2.45000) {+4};
\draw[black!12,line width=.3pt] (10.40200,0.24000) -- (10.40200,-2.19000);
\node[] at (10.40200,-2.45000) {+6};
\draw[black!12,line width=.3pt] (11.14000,0.24000) -- (11.14000,-2.19000);
\node[] at (11.14000,-2.45000) {+8};
\draw[aegisInk!70,line width=.6pt] (8.18800,0.24000) -- (8.18800,-2.19000);
\draw[aegisTeal,line width=1pt] (8.51864,0.08500) -- (9.67990,0.08500);
\draw[aegisTeal,line width=.6pt] (8.51864,0.04000) -- (8.51864,0.13000);
\draw[aegisTeal,line width=.6pt] (9.67990,0.04000) -- (9.67990,0.13000);
\path[draw=aegisTeal,fill=aegisTeal,line width=.6pt] (9.03605,0.08500) circle (0.052);
\draw[aegisInk!65,line width=1pt] (8.45960,-0.08500) -- (9.75421,-0.08500);
\draw[aegisInk!65,line width=.6pt] (8.45960,-0.13000) -- (8.45960,-0.04000);
\draw[aegisInk!65,line width=.6pt] (9.75421,-0.13000) -- (9.75421,-0.04000);
\path[draw=aegisInk!65,fill=white,line width=.6pt] (9.03605,-0.08500) circle (0.052);
\node[anchor=east] at (11.14000,-0.32000) {Holm $p=0.0249$};
\draw[aegisTeal,line width=1pt] (8.08852,-0.82500) -- (10.27867,-0.82500);
\draw[aegisTeal,line width=.6pt] (8.08852,-0.87000) -- (8.08852,-0.78000);
\draw[aegisTeal,line width=.6pt] (10.27867,-0.87000) -- (10.27867,-0.78000);
\path[draw=aegisTeal,fill=aegisTeal,line width=.6pt] (9.09575,-0.82500) circle (0.052);
\draw[aegisInk!65,line width=1pt] (8.03952,-0.99500) -- (10.35270,-0.99500);
\draw[aegisInk!65,line width=.6pt] (8.03952,-1.04000) -- (8.03952,-0.95000);
\draw[aegisInk!65,line width=.6pt] (10.35270,-1.04000) -- (10.35270,-0.95000);
\path[draw=aegisInk!65,fill=white,line width=.6pt] (9.09575,-0.99500) circle (0.052);
\node[anchor=east] at (11.14000,-1.23000) {Holm $p=0.6670$};
\draw[aegisTeal,line width=1pt] (7.88955,-1.73500) -- (9.76380,-1.73500);
\draw[aegisTeal,line width=.6pt] (7.88955,-1.78000) -- (7.88955,-1.69000);
\draw[aegisTeal,line width=.6pt] (9.76380,-1.78000) -- (9.76380,-1.69000);
\path[draw=aegisTeal,fill=aegisTeal,line width=.6pt] (8.66903,-1.73500) circle (0.052);
\draw[aegisInk!65,line width=1pt] (7.84786,-1.90500) -- (9.79043,-1.90500);
\draw[aegisInk!65,line width=.6pt] (7.84786,-1.95000) -- (7.84786,-1.86000);
\draw[aegisInk!65,line width=.6pt] (9.79043,-1.95000) -- (9.79043,-1.86000);
\path[draw=aegisInk!65,fill=white,line width=.6pt] (8.66903,-1.90500) circle (0.052);
\node[anchor=east] at (11.14000,-2.14000) {Holm $p=0.8799$};
\node[] at (9.29500,-2.80000) {AEGIS minus comparator (pp)};
\draw[aegisTeal,line width=1pt] (3.08000,-3.22000) -- (3.58000,-3.22000);
\path[draw=aegisTeal,fill=aegisTeal,line width=.6pt] (3.33000,-3.22000) circle (0.052);
\node[anchor=west] at (3.71000,-3.22000) {Project bootstrap};
\draw[aegisInk!65,line width=1pt] (7.54000,-3.22000) -- (8.04000,-3.22000);
\path[draw=aegisInk!65,fill=white,line width=.6pt] (7.79000,-3.22000) circle (0.052);
\node[anchor=west] at (8.17000,-3.22000) {Hierarchical bootstrap};
\end{tikzpicture}
 \caption{Paired F$_1$@5 and MAP differences from the primary comparators. Intervals are 95\% bootstrap confidence intervals; filled and open markers
denote project and hierarchical resampling, respectively.}
 \label{fig:paired-forest}
 \Description{Two panels show F1 at five and MAP differences for the
 foundation, equal-weight hybrid, and same-graph diffusion. Each comparison
 has two horizontal confidence intervals around the same paired mean.
 A vertical zero-reference line aligns the intervals, and adjusted p values
 are printed beside each comparison.}
\end{figure}
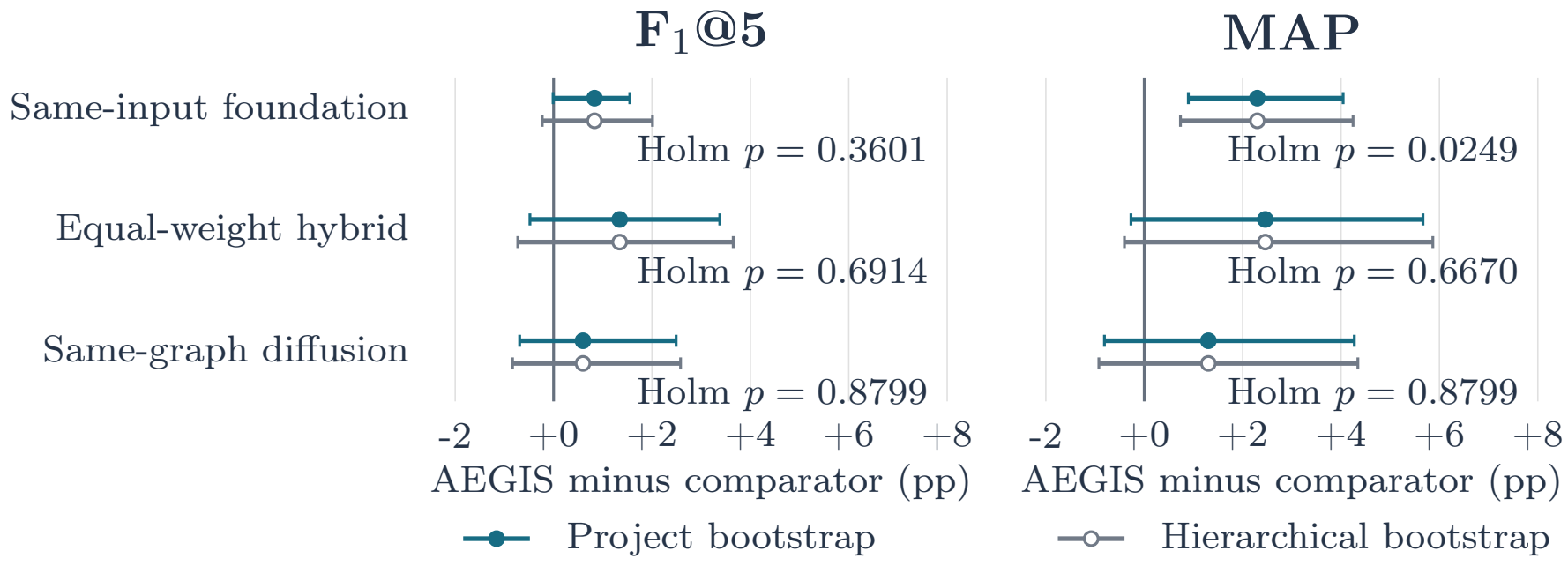

\paragraph{Project-level effects.}
MAP improves over the foundation on ten projects. All six primary
contrasts have positive macro estimates; Figure~\ref{fig:paired-forest}
shows their project and hierarchical bootstrap intervals. Both MAP
intervals against the foundation lie above zero, with
$p_H=2.49\times10^{-2}$. Supplementary Sections~S2.2--S2.3 give
project-level scores, intervals, and paired effect sizes.

All ten macro differences from the foundation remain positive when any
one project is omitted. Annotation-family weighting also preserves the
advantage over the three primary controls on all ten means
(Supplementary Section~S2.5).

\begin{rqanswer}{RQ1}
With identical retrieval inputs and per-requirement output counts, \method{}
improves native F$_1$ by 6.14\% over the retrieval foundation, with gains on
12 of 13 projects. It also improves all ten project-macro recovery and
ranking metrics over this control and leads nine of ten measures in the full
comparison. Relative F$_1$@5 gains over the same-corpus TAROT-VSM and
FTLR+MC adaptations are 33.21\% and 52.28\%, respectively.
\end{rqanswer}

\subsection{RQ2: Set Construction and Revision}
\label{sec:draft-results-decisions}
We examine retrieval evidence and reference construction, followed by
structural ordering and admission.

\paragraph{Assigning output counts to requirements.}
Compared with permuting its counts across requirements, \method{}
improves mean native F$_1$ by 13.25\%, with gains on all 13 projects
($p_H=1.95\times10^{-3}$). Permutation preserves each project's total
count and count distribution. With foundation rankings fixed, the
original counts also improve F$_1$ by 12.26\% over permutation
($p_H=1.03\times10^{-2}$). Supplementary Section~S3.4 reports the
fixed, uniform, and permuted count comparisons.

\paragraph{Building a jointly supported reference set.}
With each requirement's allocated count held fixed, the largest increase in
native-set quality occurs between the retrieval
foundation and joint reference selection; MAP also rises
(Table~\ref{tab:stages}). Admission further increases the mean native
F-measures. Witness ordering raises MAP with reference membership fixed.
\begin{table}[htbp]\centering\small
\caption{Effects of joint selection, admission, and ordering.}\label{tab:stages}
\renewcommand{\arraystretch}{1.10}\begin{tabular}{@{}lrrrr@{}}\toprule
Configuration & F$_1$@5 & MAP & Native F$_1$ & Native F$_2$ \\
\midrule
Same-input foundation & 0.2255 & 0.3424 & 0.2537 & 0.2880 \\
Joint reference selection & \textbf{0.2343} & 0.3575 & 0.2661 & 0.3008 \\
Witness order, reference set fixed & 0.2337 & 0.3650 & 0.2661 & 0.3008 \\
\method{} & 0.2338 & \textbf{0.3654} & \textbf{0.2693} & \textbf{0.3045} \\
\bottomrule\end{tabular}
% AEGIS table note begin
\par\smallskip
\begin{minipage}{\linewidth}\footnotesize\raggedright
Native metrics use the same per-requirement output count; MAP uses the complete ranking.
\end{minipage}
% AEGIS table note end
\end{table}

Across the five retrieval input pairs, retaining observed connections
in reference compatibility gives
higher final-set quality than using the degree-background penalty alone.
Both configurations use the same downstream proposal and admission
procedures. Mean native F$_1$ is 0.2275 versus 0.2218, and F$_2$ is
0.2746 versus 0.2685, with Holm-adjusted $p=0.0317$ and $p=0.0483$,
respectively. Scores are first averaged across input pairs within each
project and then across the 13 projects.

\paragraph{Retrieval evidence.}
BM25e and Jina supply complementary candidates: 30.31\% of the 8,245
gold links occur in just one view's top-50 pool. Supplementary
Section~S3.4 reports candidate coverage under single-view, union, and
matched-size RRF pooling.

\paragraph{Recovering distributed implementations.}
\label{sec:draft-structural-recovery}
Against graph-empty controls, structural processing increases recall
of links missed by both textual views on every project at depth 20,
under both Grouped--Jina and purely textual BM25e--Jina inputs.
At native counts, the primary input adds 133 and displaces 38 such links,
a net gain of 95 at the same project budget. Supplementary Section~S3.5
gives the complete counts and source-code examples.

In Infinispan, requirement \texttt{req192.txt} requests a
\texttt{TimeService} wrapping the system's time functions.
\texttt{EmptyConstructorFactory.java} creates \texttt{DefaultTimeService}
when that component is requested. BM25e and Jina rank the factory 610th
and 381st; \method{} ranks it seventh, versus 345th without the graph.
Its connections to the service interface and implementation reveal
its role in creating the service.

\paragraph{Protecting membership during revision.}
\begin{table}[htbp]
\centering\small
\caption{Recovery quality and revision downside under alternative admission rules.}
\label{tab:admission-downside}
\begin{tabularx}{\linewidth}{@{}>{\raggedright\arraybackslash}Xrrrrr@{}}\toprule
Admission & \shortstack{$\overline L$\\$(\times10^{-3})$} & Native F$_2$ & Added & Removed & Net \\ \midrule
Rank-only & 13.154 & 0.2992 & 148 & 104 & 44 \\
No core & 6.821 & 0.3028 & 113 & 58 & 55 \\
\method{} & 6.822 & 0.3045 & 134 & 64 & 70 \\
\bottomrule\end{tabularx}

% AEGIS table note begin
\par\smallskip
\begin{minipage}{\linewidth}\footnotesize\raggedright
$\overline L$: mean downside, in units of $10^{-3}$ (lower is better). F$_2$ is unscaled. Scores are project-macro means; Added/Removed/Net are total gold-link changes.
No core removes both membership protection and core-call context.
The unchanged reference set has native F$_2$ of 0.3008.
\end{minipage}
% AEGIS table note end
\end{table}
Call-context admission adds 134 gold links and removes 64, a net gain
of 70 versus 44 for rank-only admission (Table~\ref{tab:admission-downside}).
With reference membership, proposal order, core, and counts shared,
it makes fewer replacements and has higher mean native F$_2$.

Relative to rank-only admission, the estimated mean paired reduction in
revision downside is $6.332\times10^{-3}$ (95\% CI
$[2.203,11.215]\times10^{-3}$; Holm-adjusted $p=0.0586$).
Supplementary Section~S3 reports the project-level contrasts and final
replacement counts; Section~S3.1 compares exchange choices at identical
per-requirement replacement counts.

Within admission, call-context matching raises net gold-link gain from
55 to 70 and native F$_2$ from 0.3033 to 0.3045
(Table~\ref{tab:call-context-ablation}). The
specificity gate reduces gold-link removals from 84 to 64 and lowers
mean downside. The method-text guard and call fragments also reduce
removals relative to their ablations. Supplementary Section~S3 gives
the eight paired comparisons.
\begin{table}[htbp]
\centering\small
\setlength{\tabcolsep}{4pt}
\caption{Effects of call-context admission components.}
\label{tab:call-context-ablation}
\begin{tabularx}{\linewidth}{@{}>{\raggedright\arraybackslash}Xrrrrr@{}}\toprule
Configuration & Native F$_2$ & \shortstack{Downside\\$(\times10^{-3})$} & Added & Removed & Net \\ \midrule
Initial matching only & 0.3033 & 5.404 & 103 & 48 & 55 \\
No specificity gate & 0.3034 & 9.142 & 164 & 84 & 80 \\
No method-text guard & 0.3044 & 7.285 & 144 & 79 & 65 \\
Whole-method evidence & 0.3050 & 7.128 & 145 & 74 & 71 \\
\method{} & 0.3045 & 6.822 & 134 & 64 & 70 \\
\bottomrule\end{tabularx}
\par\smallskip
\begin{minipage}{\linewidth}\footnotesize\raggedright
Native F$_2$ and downside are project-macro means; link changes are totals across the 13 projects. Downside and link changes are measured from the original reference sets. Initial matching only omits the call-context matching stage. The other controls each change one component of that stage.
\end{minipage}
\end{table}

\begin{rqanswer}{RQ2}
At fixed per-requirement counts, observed code connections in reference
construction improve final native F$_1$ and F$_2$ over degree-only
compatibility. Retrieval views supply complementary candidates, and witness
ordering raises mean MAP without changing reference membership.
Under the default input, admission adds
134 gold links and removes 64, giving a net gain of 70 without changing
output counts.
\end{rqanswer}

\subsection{RQ3: Retrieval Inputs and Efficiency}
\label{sec:draft-results-robustness}
We examine recovery across retrieval inputs and parameter settings,
and execution cost once retrieval inputs are available.

\paragraph{Recovery across retrieval foundations.}
With the same construction and admission rules, \method{} improves
project-macro native F$_1$, native F$_2$, F$_1$@5, and MAP over the
corresponding retrieval foundation for all five input pairs
(Table~\ref{tab:foundation-recovery}). The inputs include grouped
retrieval, a single BM25-based lexical view, alternative semantic
encoders, and the entirely lexical BM25e--VSM pair.
\begin{table}[htbp]
\centering\small\setlength{\tabcolsep}{3pt}
\caption{Recovery across retrieval foundations.}
\label{tab:foundation-recovery}
\begin{tabularx}{\linewidth}{@{}>{\raggedright\arraybackslash}Xlrrrr@{}}\toprule
Retrieval input & Output & Native F$_1$ & Native F$_2$ & F$_1$@5 & MAP \\ \midrule
Grouped--\allowbreak Jina & Foundation & 0.2537 & 0.2880 & 0.2255 & 0.3424 \\
& \method{} & \textbf{0.2693} & \textbf{0.3045} & \textbf{0.2338} & \textbf{0.3654} \\
\addlinespace[2pt]
Grouped--\allowbreak UniXcoder & Foundation & 0.2122 & 0.2452 & 0.1925 & 0.2907 \\
& \method{} & \textbf{0.2260} & \textbf{0.2587} & \textbf{0.2062} & \textbf{0.3100} \\
\addlinespace[2pt]
BM25e--\allowbreak Jina & Foundation & 0.2243 & 0.2880 & 0.2217 & 0.3383 \\
& \method{} & \textbf{0.2413} & \textbf{0.3087} & \textbf{0.2280} & \textbf{0.3525} \\
\addlinespace[2pt]
BM25e--\allowbreak UniXcoder & Foundation & 0.1751 & 0.2323 & 0.1704 & 0.2635 \\
& \method{} & \textbf{0.1913} & \textbf{0.2501} & \textbf{0.1832} & \textbf{0.2812} \\
\addlinespace[2pt]
BM25e--\allowbreak VSM & Foundation & 0.2012 & 0.2396 & 0.1831 & 0.2737 \\
& \method{} & \textbf{0.2098} & \textbf{0.2507} & \textbf{0.1902} & \textbf{0.2923} \\
\bottomrule\end{tabularx}
\par\smallskip
\begin{minipage}{\linewidth}\footnotesize\raggedright
Scores are macro-averaged over 13 projects. Native comparisons use
matched output counts within each input pair. Bold indicates the
higher mean within each comparison.
\end{minipage}
\end{table}

Averaging scores across input pairs within each project gives relative gains
of 6.67\% in native F$_1$ and 6.17\% in F$_2$. Both gains are positive on
all 13 projects, with $p_H=2.93\times10^{-3}$ for each metric. Supplementary
Section~S3.2 gives the full comparisons.

\paragraph{Sensitivity of set construction and revision.}
Raising the budget threshold from 0.50 to 0.70 reduces mean output
count from 11.96 to 3.42; the default 0.60 gives 6.07. Across the
12 core-fraction--tolerance settings, native F$_1$ ranges from
0.2656 to 0.2693 and MAP from 0.3648 to 0.3657. Mean downside rises
with tolerance at each core fraction; protecting 0.75 of the reference
set gives lower downside than 0.25 or 0.50 at every tolerance. Changing
the core also changes available call context. Increasing specificity
threshold $\alpha$ from 0.01 to 0.20 raises final entrant counts and
downside, with F$_1$ within 0.2686--0.2697. Supplementary Section~S3.3
reports all parameter settings and results.

\paragraph{Execution cost.}
With retrieval inputs available, the measured times for set construction
at all sizes, budget allocation, and reference-priority construction sum
to 22.945 seconds across the 13 projects. Preparing source-fragment scores
from parsed-call records takes 30.491 seconds. With source features and
witness scores available, median admission time ranges from 1.38 to
362.66\,ms per project over three repetitions; the per-project medians
sum to 0.792 seconds. Supplementary Section~S2.4 reports the
project-level timings, and Section~\ref{sec:draft-experiments} specifies
the measurement procedure.

\begin{rqanswer}{RQ3}
With scores averaged across the five retrieval pairs within projects,
\method{}
improves native F$_1$ by 6.67\% and F$_2$ by 6.17\% at matched counts,
with both gains positive on all 13 projects. These pairs include purely
lexical inputs. With source features and witness scores available,
admission takes 1.38--362.66\,ms per project.
\end{rqanswer}

\section{Discussion}
\label{sec:draft-discussion}
The allocation of review positions and the artifacts filling them both
affect recovery.

Permuting the allocated counts preserves the project total and count
distribution; the higher F$_1$ under the original allocation shows that
assigning these counts to individual requirements affects recovery.
The artifacts filling those positions also matter. At matched counts,
the complete pipeline improves recovery over the retrieval foundation,
and observed connections contribute beyond the degree-background penalty.

A structural candidate's value depends on the member it would displace.
Retrieval evidence and calls involving protected members can justify
replacing one member while retaining another with stronger evidence.
Only unprotected positions change, and each requirement's count stays fixed.

Developers can use the selected sets and their inspection order to locate
implementation code. A team can prepare retrieval scores, code relations,
and call fragments from requirements and source code without existing
project trace-link labels. For each changed position, \method{} records
the original and final members, structural support, retrieval-score
differences, and supporting calls. A review interface could display
these records alongside the proposed links.

\section{Threats to Validity}
\label{sec:draft-threats}
We discuss threats to the findings on trace-set quality at matched output
counts and on changes made during revision.

\subsection{Construct Validity}
Matching project budgets controls the total number of candidate links
returned for inspection. Matching each requirement's count also fixes
the distribution of those positions. Actual review effort depends on
artifact size, complexity, and developer familiarity.

The trace matrices contain requirement-to-class, use-case-to-class, and
issue/change-derived relations, with different annotation conventions
for indirect participation. Recovery scores measure agreement with these
links, so valid links absent from the matrices count as false positives.
The linked-requirement analysis checks the effect of requirements without
aligned positive links.

The admission bound limits normalized retrieval-score loss, while
revision downside measures the decrease in native-set F$_2$ from the
reference set against the supplied links. We report final-set quality
and the numbers of gold links added and removed alongside downside.

\subsection{Internal Validity}
Controls fix retrieval inputs, counts, or reference sets according to
the tested decision. Matched-activity controls also fix replacement counts. The graph-empty
comparison uses purely textual retrieval inputs, while compatibility
controls hold pools, counts, and downstream procedures fixed. Supplementary Section~S4.4 documents preprocessing, parsing
fallbacks, and structural substitutions in the TAROT, FTLR, and O-CSTI
adaptations. Numerical checks compare adapted TAROT and FTLR operations
with the original implementations.

\subsection{External Validity}
The evaluation covers thirteen Java projects from several domains and
with different candidate-set sizes. Alternative-input experiments and
sensitivity analyses examine robustness across retrieval inputs and
parameter settings within this collection.

\method{} depends on the relevance evidence and code relationships
available in a project. Five projects use call-derived relations for the
proposal graph, while eight use import and inheritance proxies.
Admission extracts source calls separately in every project; unresolved
calls and empty fragments provide no context support. The alternative-input
experiments include a purely lexical retrieval pair.

A developer study would assess how the exchange records affect
maintenance-task performance.

\subsection{Conclusion Validity}
Requirements within a project share code artifacts and may repeat
annotated link sets, so we use projects as the paired units of inference.
The resampling and sign-flip procedures rely on the exchangeability
assumptions stated in the supplementary protocol.

\section{Related Work}
\label{sec:draft-related-work}
Traceability reviews cover retrieval techniques and their use in
development~\cite{borg2014mapping,wang2018review}. We focus on retrieval,
structural evidence, and candidate selection or validation.

\subsection{Retrieval Representations and Fusion}
Early IR approaches match documentation with source
code~\cite{antoniol2002recovering}. Marcus and
Maletic~\citeyearpar{marcus2003lsi} use LSI to connect artifacts beyond
exact term overlap; De Lucia et al.~\citeyearpar{delucia2007recovering}
integrate LSI into artifact management and incremental recovery.
Hayes et al.~\citeyearpar{hayes2006candidate} study candidate generation
and analyst feedback. Gethers et al.~\citeyearpar{gethers2011orthogonal}
combine vector-space, probabilistic, and topic models to recover links
missed by individual representations. \method{} uses complementary
retrieval scores in a set objective that also considers structural compatibility.

FTLR aggregates Word Mover's Distance-based sentence-to-method matches
into class links through voting, with variants using comments and call
dependencies~\cite{hey2021ftlr}. NoRBERT filters requirement fragments
by aspects and concerns before FTLR recovery~\cite{hey2024requirementsclassification}.
TAROT enriches text and adjusts IR similarity with consensual
biterms~\cite{gao2022tarot}; AVIATE uses biterms shared across translation
variants in bilingual projects~\cite{sun2024aviate}. CodeBERT, UniXcoder,
and Jina support semantic matching between language and
code~\cite{feng2020codebert,guo2022unixcoder,kryvosheieva2025jina}.
These methods change representations or aggregate textual matches.
\method{} applies set-construction and revision rules to the resulting scores.

\subsection{Structural and Transitive Recovery}
McMillan et al.~\citeyearpar{mcmillan2009structural} combine textual and
structural analysis to recover links between documentation artifacts.
Panichella et al.~\citeyearpar{panichella2013structural}
regulate structural enhancement with developer feedback; Kuang et
al.~\citeyearpar{kuang2019frugal} combine limited feedback with call and
data-dependency closeness. \method{} constructs reference sets without
confirmed links as input, then assesses structural candidates using
retrieval and call-context evidence. Ghabi and
Egyed~\citeyearpar{ghabi2012patterns} validate recorded traces against
calling patterns; our admission test assesses proposed changes to retrieved sets.

TRIAD uses intermediate artifacts and transitive links to enrich text
and adjust similarity~\cite{gao2024triad}. TransArC bridges architecture
documentation and code through a component model, combining ArDoCo
and ArCoTL~\cite{keim2024transarc}. \method{} draws proposal support
from directed code paths and multiple ranked code anchors.

\subsection{Learned and Generative Link Prediction}
COMET estimates link probabilities from textual similarities, feedback,
and transitive relations through a hierarchical Bayesian
model~\cite{moran2020comet}. Guo et al.~\citeyearpar{guo2017semantic}
learn artifact relations with embeddings and recurrent networks;
HGNNLink combines pretrained representations, IR similarity, and code
dependencies in a heterogeneous graph network~\cite{wang2025hgnnlink}.
\method{} uses fixed scores and code evidence without training a
trace-link predictor on project labels.

T-BERT transfers pretrained representations to trace prediction,
including issue-to-commit recovery~\cite{lin2021tbert}. EALink uses
distillation and contrastive learning~\cite{zhang2023ealink}; SPLINT
adds semi-supervised preprocessing~\cite{dong2022splint}, and DSSLink
expands training with pseudo-labels~\cite{zhu2024dsslink}. These
issue-to-commit applications concern change history; our task selects
implementation artifacts from a supplied code collection.

LiSSA uses retrieval-augmented generation for mapping decisions across
requirements, code, and architecture artifacts~\cite{fuchss2025lissa}.
DSDR combines semantic-gated diffusion and backward verification,
then passes evidence paths to an LLM validator~\cite{zhao2026dsdr}.
Our admission rule compares a candidate with its possible donor using
retrieval and call context without an LLM validator.

\subsection{Set Selection and Inspection Budgets}
Maximal marginal relevance balances query relevance against similarity
to selected documents to reduce redundancy~\cite{carbonell1998mmr}.
Our pairwise term instead rewards structural support above a graph
background and penalizes support below it. \method{} also compares
sets at different sizes when allocating a project budget.
ENRL estimates positive links remaining in ranked lists with
classifiers~\cite{falessi2017remaining}. \method{} allocates positions
by candidate-set scores; subsequent revision preserves these sizes
and bounds retrieval-evidence loss.

\subsection{Trace Explanations and Maintenance}
Liu et al.~\citeyearpar{liu2022explanations} explain candidate links
through domain concepts, definitions, and artifact relations, and
evaluate a visual interface. Our exchange records identify the added and
removed artifacts and the evidence supporting each replacement.
TRAIL learns trace-maintenance decisions from existing
traces~\cite{mills2018trail}; FeatRacer learns feature-recording practices
from developer annotations to recommend missing annotations~\cite{mukelabai2023featracer}. \method{} recovers sets in one run
from retrieval scores, code relations, and call fragments without
existing trace links as input.

\section{Conclusion}
\label{sec:draft-conclusion}
This work studied how retrieval evidence and code relations can improve
requirement-to-code trace sets under a limited review budget. \method{}
uses requirement relevance and structural compatibility to allocate review
positions across requirements and select the artifacts that fill them.
Call-context admission then compares structural candidates with the members
they would replace. Accepted replacements preserve the allocated set sizes
and protected members and satisfy a bound on retrieval-score loss.

Across 13 projects, \method{} achieved the highest project-macro average
on nine of ten recovery and ranking metrics among the compared methods.
Its F$_1$@5 exceeded the same-corpus adaptations of TAROT-VSM and FTLR+MC
by 33.21\% and 52.28\%, respectively. Recovery also improved over the
corresponding retrieval foundations across five input pairs at matched
per-requirement output counts. Controlled comparisons showed that observed
code connections contributed beyond degree-based penalties, and admission
further increased the number of gold links in the selected sets without
increasing their sizes.
\method{} also records the evidence behind membership changes, allowing
developers to inspect the returned artifacts and the reasons for
each replacement.

\section*{Acknowledgment}
This work was supported by JST SPRING, Grant Number JPMJSP2138.

\section*{Declaration of generative AI and AI-assisted technologies in the manuscript preparation process}
The authors used ChatGPT to improve the manuscript's language, clarity,
and readability. The authors reviewed the suggestions, revised the text,
and take responsibility for the final manuscript.

\bibliographystyle{elsarticle-harv}
\bibliography{aegis/references-aegis,aegis/references-aegis-showcase-extra,aegis/references-aegis-foundations}
\end{document}